\documentclass[prb,preprint]{revtex4}
\usepackage{graphicx}
\begin{document}
\title{Doping Dependence of the Redistribution of Optical Spectral Weight in Bi$_{2}$Sr$_{2}$CaCu$_{2}$O$_{8+\delta}$}
\author{F. Carbone, A.B. Kuzmenko,  H.J.A. Molegraaf,
E. van Heumen, V. Lukovac, F. Marsiglio and D. van der Marel}
\affiliation{ Departement de Physique de la Mati\`{e}re
Condens\'{e}e, Universit\'{e} de Gen\`{e}ve, 24 Quai
Ernest-Ansermet, CH-1211 Geneva 4, Switzerland \\}
\author{K. Haule, G. Kotliar}
\affiliation{Departement of Physics, Rutgers University, Piscataway,
NJ 08854, USA\\}
\author{H. Berger,S. Courjault}
\affiliation{\'{E}cole Polytechnique Federale de Lausanne,
Departement de Physique, CH-1015 Lausanne, Switzerland \\}

\author{P.H. Kes,M. Li}
\affiliation{Kamerlingh Onnes Laboratory, Leiden University, 2300 RA
Leiden, The Netherlands \\}

\date{\today}

\begin{abstract}

We present the ab-plane optical conductivity of four single crystals
of Bi$_{2}$Sr$_{2}$CaCu$_{2}$O$_{8+\delta}$ (Bi2212) with different
carrier doping levels from the strongly underdoped to the strongly
overdoped range with $T_c$=66, 88, 77, and 67 K respectively. We
focus on the redistribution of the low frequency optical spectral
weight (SW) in the superconducting and normal states. The
temperature dependence of the low-frequency spectral weight in the
normal state is significantly stronger in the overdoped regime. In
agreement with other studies, the superconducting order is marked by
an increase of the low frequency SW for low doping, while the SW
decreases for the highly overdoped sample. The effect crosses
through zero at a doping concentration $\delta$=0.19 which is
slightly to the right of the maximum of the superconducting dome.
This sign change is not reproduced by the BCS model calculations,
assuming the electron-momentum dispersion known from published ARPES
data. Recent Cluster Dynamical Mean Field Theory (CDMFT)
calculations based on the Hubbard and t-J models, agree in several
relevant respects with the experimental data.

\end{abstract}

\maketitle

\section{INTRODUCTION}\label{intro}

One of the most puzzling phenomena in the field of high temperature
superconductivity is the doping dependence of the electronic
structure of the cuprates. Several experiments report a conventional
Fermi Liquid behavior on the overdoped side of the superconducting
'dome' \cite{Proust,arpes1,specheat,deutscher}, while the enigmatic
'pseudogap phase' is found in underdoped
samples\cite{NMR,ARPES,optics}. In the underdoped and optimally
doped regions of the phase diagram it has been shown for bi-layer
Bi2212 \cite{hajo,santander,comment} and tri-layer Bi2223
(Bi$_{2}$Sr$_{2}$Ca$_{2}$Cu$_{3}$O$_{10}$) \cite{io} that the
superconductivity induced low frequency Spectral Weight (SW)
increases when the system becomes superconducting. This observation
points toward a non BCS-like pairing mechanism, since in a BCS
scenario the superconductivity induced SW transfer would have the
opposite sign. On the other hand, in Ref. \onlinecite{deutscher} a
fingerprint of more conventional behavior has been reported using
optical techniques for a strongly overdoped thin film of Bi2212: the
SW redistribution at high doping has the opposite sign with respect
to the observation for under and optimal doping.

It is possible to relate the SW transfer and the electronic kinetic
energy using the expression for the intraband spectral weight $W$
via the energy momentum dispersion n$_k$ of the conduction electrons
\cite{kinsw}
\begin{equation}\label{equation1}
    W(\Omega_c,T) \equiv {\int}^{\Omega_c}_0 \sigma_1(\omega,T)d\omega =
     \frac{\pi e^2 a^2}{2\hbar^2V}<-\hat{K}>,
\end{equation}
\noindent where $\sigma_1(\omega,T)$ is the real part of the optical
conductivity, $\Omega_c$ is a cutoff frequency, $a$ is the in-plane
lattice constant, $V$ is the volume of the unit cell and
$\hat{K}\equiv -a^{-2}\sum_k \hat{n}_k
\partial^2\epsilon_k/\partial k^2$. The operator $\hat{K}$ becomes the exact kinetic
energy $\sum_k \hat{n}_k \epsilon_k$ of the free carriers within the
nearest neighbor tight-binding approximation. It has been shown, in
Refs. \onlinecite{dirk, frank}, that even after accounting for the
next nearest neighbor hopping parameter the exact kinetic energy and
$<-\hat{K}>$ approximately coincide and follow the same trends as a
function of temperature. According to Eqn. (1), the lowering of
$W(\Omega_c)$ implies an increase of the electronic kinetic energy
and vice-versa. In this simple scenario a decrease of the low
frequency SW, when the system becomes superconducting, would imply a
superconductivity induced increase of the electronic kinetic energy,
as it is the case for BCS superconductors.

In the presence of strong electronic correlations this basic picture
has to be extended to take into account that at different energy
scales materials are described by different model Hamiltonians, and
different operators to describe the electric current at a given
energy scale \cite{eskes, rozenberg}. In the context of the Hubbard
model, Wrobel {\em et al.} pointed out\cite{wrobel} that if the
cutoff frequency $\Omega_c$ is set between the value of the exchange
interaction $J \simeq 0.1$ eV and the hopping parameter $t \simeq
0.4$ eV then $W(\Omega_c)$ is representative of the kinetic energy
of the holes within the t-J model in the spin polaron approximation
and describes the excitations below the on-site Coulomb integral $U
\simeq 2$ eV not involving double occupancy, while $W(\Omega_c > U)$
represents all intraband excitations and therefore describes the
kinetic energy of the full Hubbard Hamiltonian. A numerical
investigation of the Hubbard model within the dynamical cluster
approximation\cite{meier} has shown the lowering of the full kinetic
energy below $T_c$, for different doping levels, including the
strongly overdoped regime. Experimentally, this result should be
compared with the integrated spectral weight where the cutoff
frequency is set well above $U = 2$ eV in order to catch all the
transitions into the Hubbard bands. However, in the cuprates this
region also contains interband transitions, which would make the
comparison rather ambiguous.

Using Cluster Dynamical Mean Field Theory (CDMFT) on a 2$\times$2
cluster Haule and Kotliar \cite{kotliar} recently found that, while
the total kinetic energy decreases below $T_c$ at all doping levels,
the kinetic energy of the holes exhibits the opposite behavior on
the two sides of the superconducting dome: In the underdoped and
optimally doped samples the kinetic energy of the holes, which is
the kinetic energy of the t-J model, increases below $T_c$. In
contrast, on the overdoped side the same quantity decreases when the
superconducting order is switched on in the calculation. This is in
agreement with the observations of Ref.\onlinecite{deutscher} as
well as the experimental data in the present paper. The good
agreement between experiment and theory in this respect is
encouraging, and it suggests that the t-J model captures the
essential ingredients, needed to describe the low energy excitations
in the cuprates, as well as the phenomenon of superconductivity
itself.

%\subsection{}\label{sec3}
%
The Hubbard model and the t-J model are based on the assumption that
strong electron electron correlations rule the physics of these
materials. Based on these models an increase of the low frequency SW
in the superconducting state was found in the limit of low doping
\cite{wrobel} in agreement with the experimental results
\cite{hajo,io}. The optical conductivity of the t-J model in a
region of intermediate temperatures and doping near the top of the
superconducting dome has been recently studied using CDMFT
\cite{kotliar}. The CDMFT solution of the t-J model at different
doping levels suggests a possible explanation for the fact that the
optical spectral weight shows opposite temperature dependence for
the underdoped and the overdoped samples. It is useful to think of
the kinetic energy operator of the hubbard model, at large U as
composed of two physically distinct contributions representing the
superexchange energy of the spins and the kinetic energy of the
holes. The superexchange energy of the spins is the result of the
virtual transitions across the charge transfer gap, thus, the
optical spectral weight integrated up to an energy below these
excitations is representative only of the kinetic energy of the
holes. The latter contribution to the total kinetic energy was found
to decrease in the underdoped regime while it increases above
optimal doping, as observed experimentally. This kinetic energy
lowering is however rather small compared to the lowering of the
superexchange energy.
%This larger energy scale is most likely responsible for the high
%T$_c$ of these materials \cite{philips}.
Upon overdoping the kinetic energy of the holes increases in the
superconducting state, while the larger decrease of the
super-exchange energy makes superconductivity favorable with a still
high value of $T_c$. In the CDMFT study of the t-J model, a stronger
temperature dependence of $W(T)$ is found in the overdoped side.
This reflects the increase in Fermi Liquid coherence with reducing
temperature.

In the present paper we extend earlier experimental studies of the
temperature dependent optical spectral weight of Bi2212 by the same
group\cite{hajo,comment} to the overdoped side of the phase diagram,
{\em i.e.}with superconducting phase transition temperatures of 77 K
and 67 K. We report a strong change in magnitude of the temperature
dependence in the normal state for the sample with the highest hole
doping, and we show that the kink in the temperature dependence at
$T_c$ changes sign at a doping level of about 19 percent, in
qualitative agreement with the report by Deutscher et al.
\cite{deutscher}.

\section{EXPERIMENT AND RESULTS}\label{sec1}

In this paper we concentrate on the properties of single crystals of
Bi2212 at 4 different doping levels, characterized by their
superconducting transition temperatures. The preparation and
characterization of the underdoped sample (UD66K), an optimally
doped crystal (Opt88) and an overdoped sample (OD77) with $T_c$'s of
66, 88 K and 77 K respectively, have been given in Ref.
\onlinecite{hajo}. The crystal with the highest doping level (OD67)
has a $T_c$ of 67 K. This sample has been prepared with the
self-flux method. The oxygen stoichiometry of the single crystal has
been obtained in a PARR autoclave by annealing for 4 days in Oxygen
at 140 atmospheres and slowly cooling from 400 °C to 100 °C. The
infrared optical spectra and the spectral weight analysis of samples
UD66 and OpD88 have been published in Refs.
\onlinecite{hajo,comment}. The phase of $\sigma(\omega)$ of sample
OD77 has been presented as a function of frequency in a previous
publication\cite{MarelNature03}. In the present manuscript we
present the optical conductivity of samples OD77 and OD67 for a
dense sampling of temperatures, and we use this information to
calculate $W(\Omega_c,T)$. The samples are large ($4\times4\times
0.2$ mm$^{3}$) single crystals. The crystals were cleaved within
minutes before being inserted into the optical cryostat. We measured
the real and imaginary part of the dielectric function with
spectroscopic ellipsometry in the frequency range between 6000 and
36000 cm$^{-1}$ (0.75 - 4.5 eV). Since the ellipsometric measurement
is done at a finite angle of incidence (in our case 74$^\circ$), the
measured pseudo-dielectric function corresponds to a combination of
the ab-plane and c-axis components of the dielectric tensor. From
the experimental pseudo-dielectric function and the published c-axis
dielectric function of Bi-2212 \cite{tajima} we calculated the
ab-plane dielectric function. In accordance with earlier results on
the cuprates \cite{bozovic,io} and with the analysis of
Aspnes\cite{aspnes}, the resulting ab-plane dielectric function
turns out to be very weakly sensitive to the c-axis response and its
temperature dependence. In the range from 100 to 7000 cm$^{-1}$
(12.5 - 870 meV) we measured the normal incidence reflectivity,
using gold evaporated {\em in situ} on the crystal surface as a
reference.
\begin{figure}[ht]
   \centerline{\includegraphics[width=8cm,clip=true]{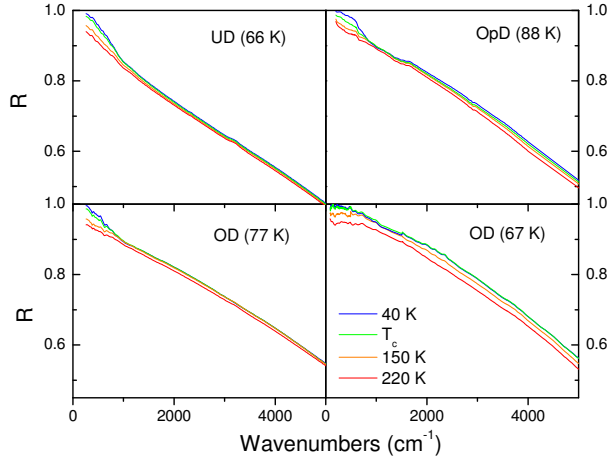}}
   \caption{Reflectivity spectra of Bi2212 at selected temperatures for different doping levels, described in the text.}
   \label{fig1}
\end{figure}

The infrared reflectivity is displayed for all the studied doping
levels in Fig. \ref{fig1}. The absolute reflectivity increases with
increasing doping, as expected since the system becomes more
metallic. Interestingly, the curvature of the spectrum also changes
from under to overdoping; this is reflected in the frequency
dependent scattering rate as has been pointed out recently by Wang
{\em et al.} \cite{TimuskNature}. In order to obtain the optical
conductivity in the infrared region we used a variational routine
that simultaneously fits the reflectivity and ellipsometric data
yielding a Kramers-Kronig (KK) consistent dielectric function which
reproduces all the fine features of the measured spectra. The
details of this approach are described elsewhere
\cite{Kuzmenko04,io}. All data were acquired in a mode of continuous
temperature scans between 20 K and 300 K with a resolution of 1 K.
Very stable measuring conditions are needed to observe changes in
the optical constants smaller than 1\%. We use home-made cryostats
of a special design, providing a temperature independent and
reproducible optical alignment of the samples. To avoid spurious
temperature dependencies due to adsorbed gases at the sample
surface, we use a Ultra High Vacuum UHV cryostat for the
ellipsometry in the visible range, operating at a pressure in the
$10^{-10}$ mbar range, and a high vacuum cryostat for the normal
incidence reflectivity measurements in the infrared, operating in
the $10^{-7}$ mbar range.

\begin{figure}[ht]
   \centerline{\includegraphics[width=8cm,clip=true]{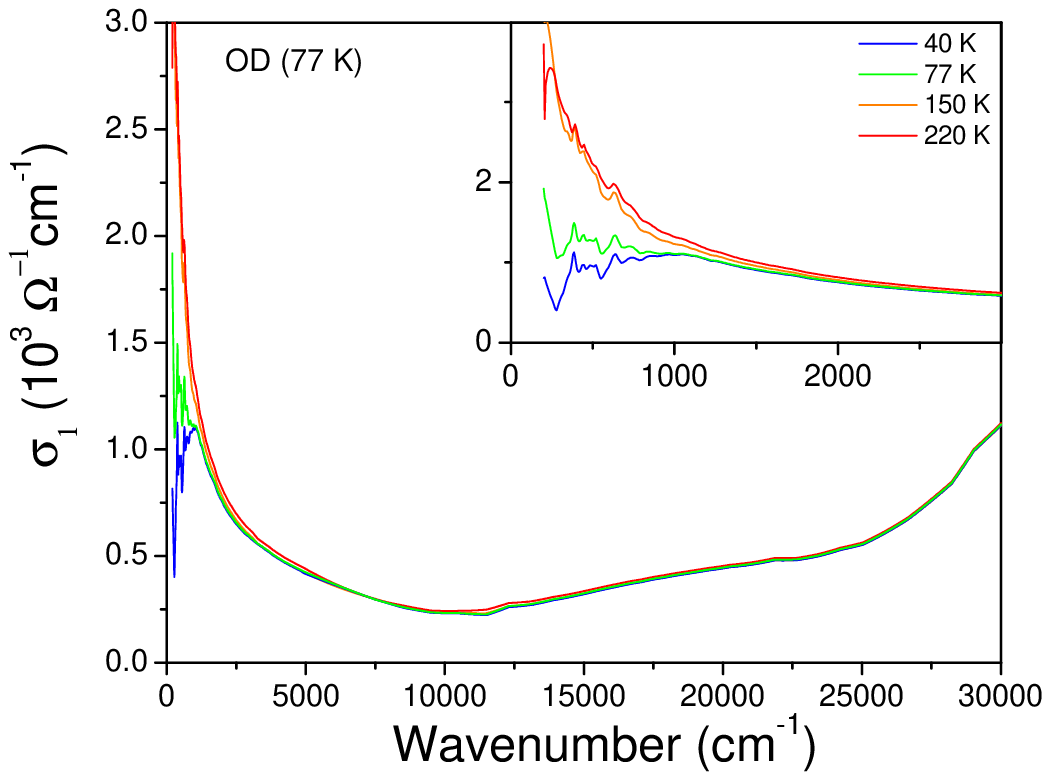}
   \includegraphics[width=8cm,clip=true]{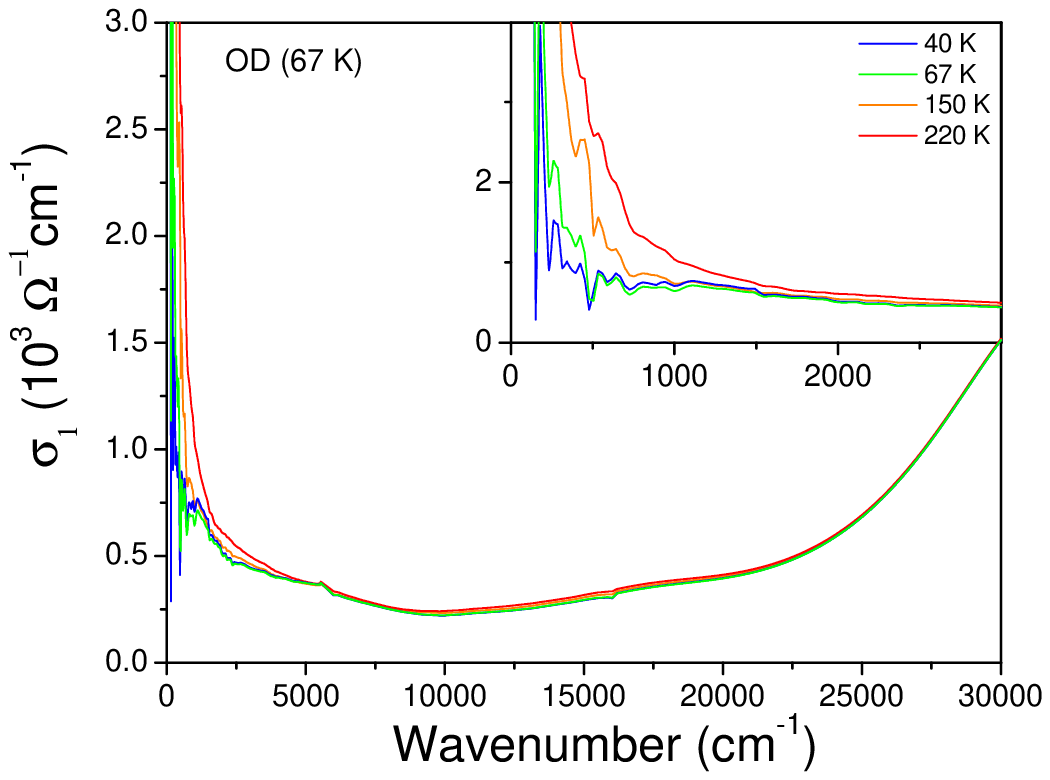}}
   \caption{In-plane optical conductivity of slightly overdoped ($T_c$= 77 K, left panel) and
   strongly overdoped ($T_c$= 67 K, right panel) samples of Bi2212 at selected temperatures.
   The insets show the low energy parts of the spectra.}
   \label{fig2}
\end{figure}

In Fig. \ref{fig2} we show the optical conductivity of the two
overdoped samples of Bi2212 with $T_c$ = 77 K and $T_c$ = 67 K at
selected temperatures. Below 700 cm$^{-1}$ one can clearly see the
depletion of the optical conductivity in the region of the gap at
low temperatures (shown in the inset). The much smaller absolute
conductivity changes at higher energies, which are not discernible
at this scale, will be considered in detail below.

\begin{figure}[ht]
    \centerline{\includegraphics[width=9cm,clip=true]{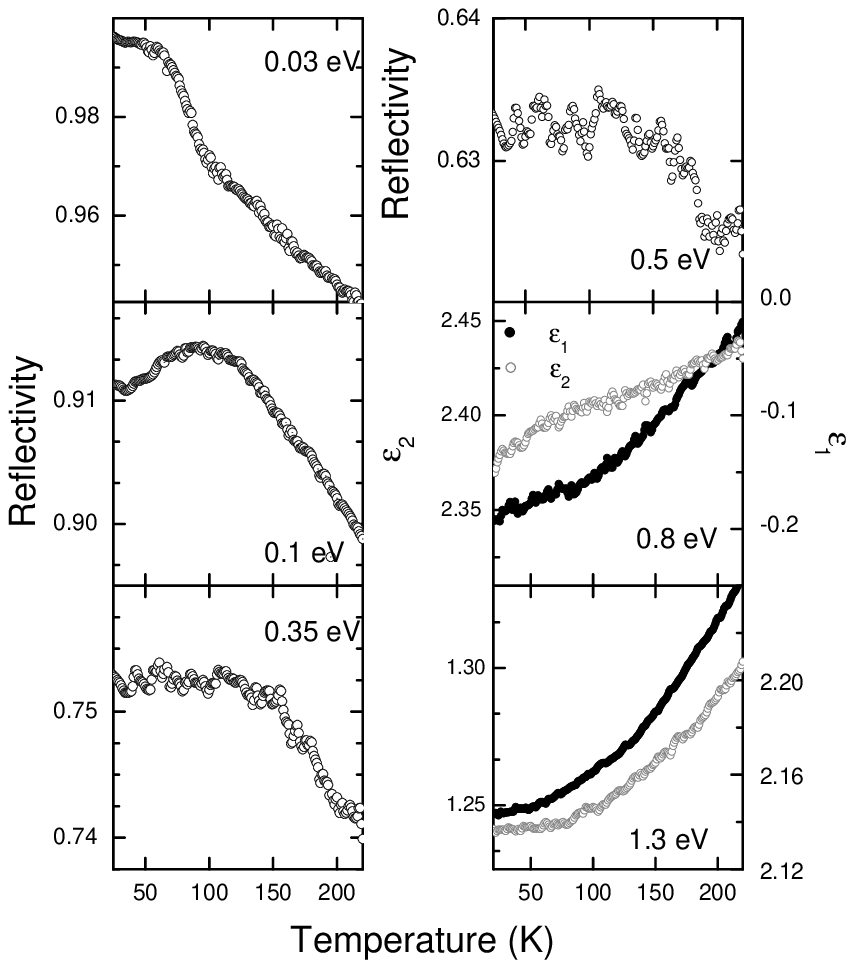}
    \includegraphics[width=9cm,clip=true]{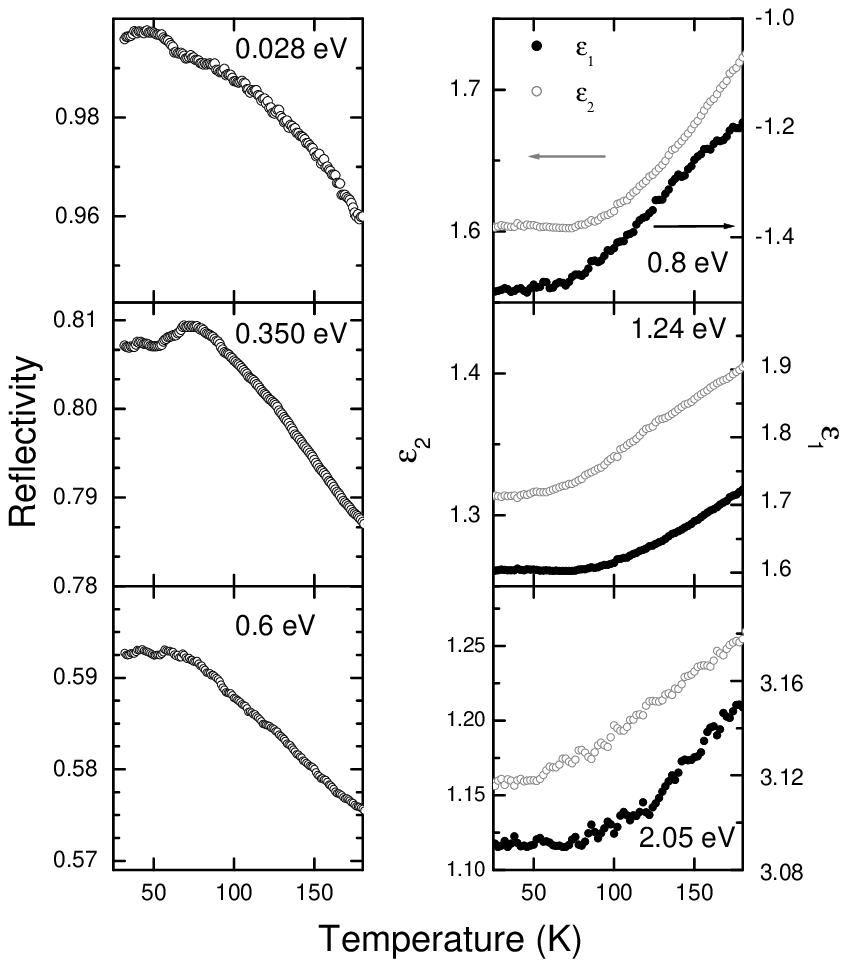}}
    \caption{Leftmost (rightmost) two columns: reflectivity and
    dielectric function of sample OD77 (OD67)
    as a function of temperature for selected photon energies.
    The corresponding photon energies are indicated in the panels.
    The real (imaginary) parts of $\epsilon(\omega)$ are indicated
    as closed (open) symbols.}
    \label{fig4}
\end{figure}

One can see the effect of superconductivity on the optical constants
in the temperature dependent traces, displayed in Fig. \ref{fig4},
at selected energies, for the two overdoped samples. In comparison
to the underdoped and optimally doped samples \cite{hajo,io} where
reflectivity is found to have a further increase in the
superconducting state at energies between 0.25 and 0.7 eV, in the
overdoped samples reflectivity decreases below $T_c$ or remain more
or less constant. In the strongly overdoped sample one can clearly
see, for example at 1.24 eV, that at low temperature $\epsilon_1$
increases cooling down, opposite to the observation on the optimally
and underdoped samples. These details of the temperature dependence
of the optical constants influence the integrated SW trend as we
will discuss later in the text.

\section{DISCUSSION}\label{sec2}
\subsection{Spectral weight analysis of the experimental data}

\begin{figure}[ht]
    \centerline{\includegraphics[width=9cm,clip=true]{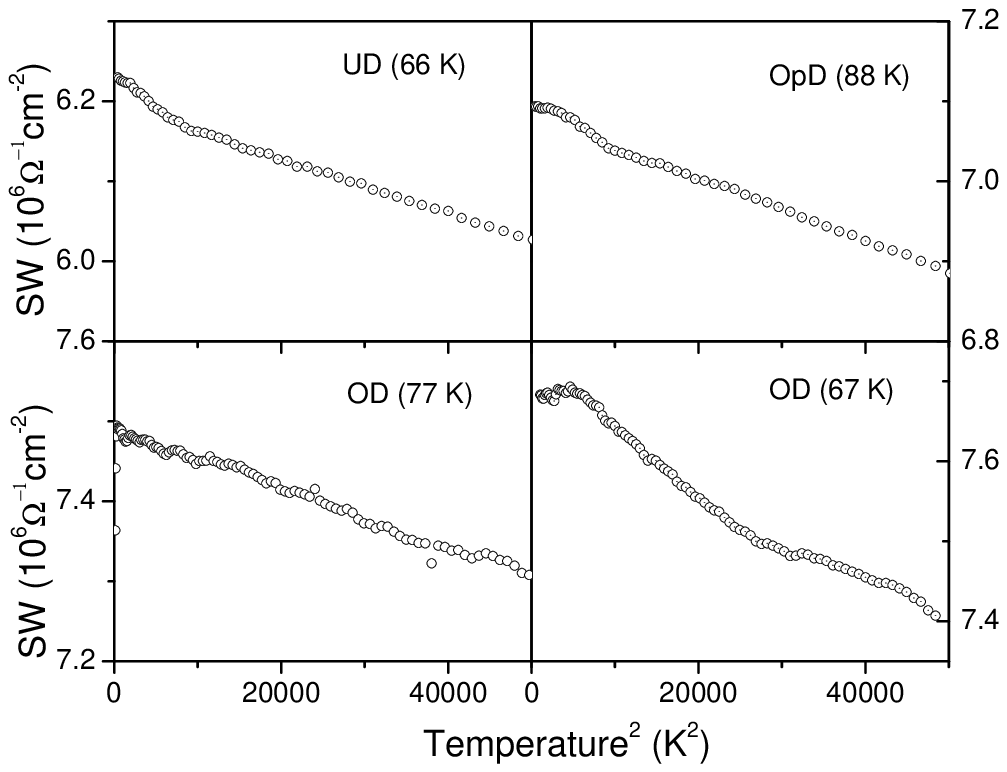}
    \includegraphics[width=9cm,clip=true]{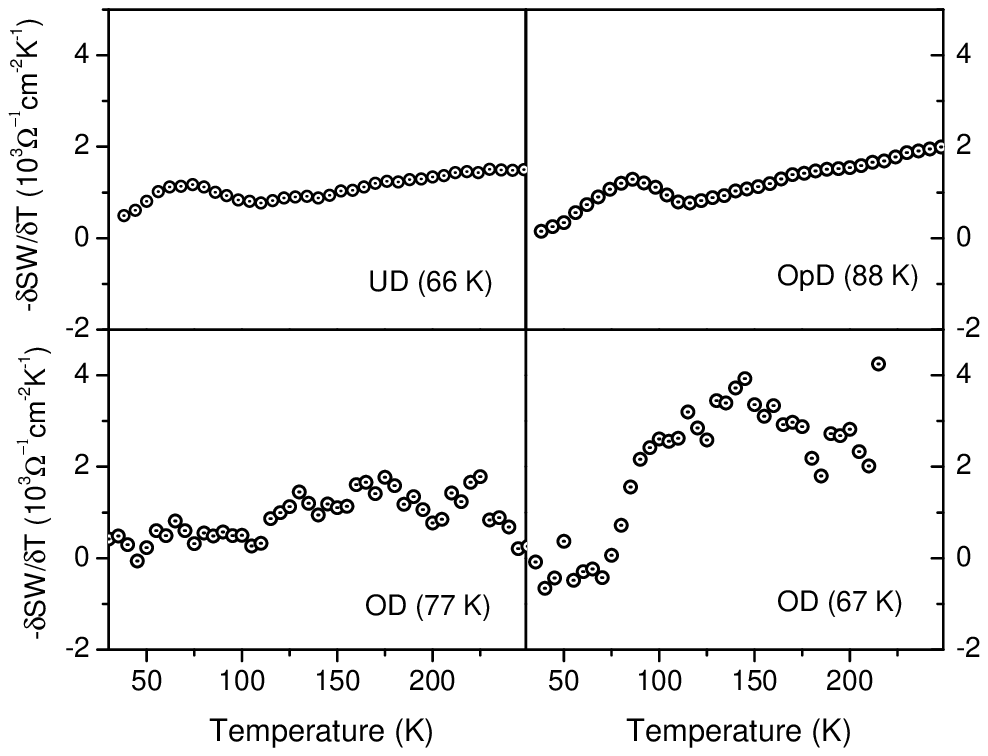}}
    \caption{Left panel: spectral weight $W(\Omega_c,T)$ for $\Omega_c$ = 1.24 eV, as a function of temperature for different doping levels.
    Right panel: the derivative ($\frac{-\partial W(\Omega_c,T)}{\partial T}$) as a function of temperature for different doping levels.
    For the derivative curves the data have been averaged in 5 K intervals in order to reduce the noise.}
    \label{fig7}
 \end{figure}

As it is discussed in our previous publications
\cite{Kuzmenko04,io,comment}, using the knowledge of both $\sigma_1$
and $\epsilon_1$ we can calculate the low frequency SW without the
need of the knowledge of $\sigma_1$ below the lowest measured
frequency. When the frequency cut off of the integral is chosen to
be lower than the charge transfer energy (around 1.5 eV), the SW is
representative of the free carrier kinetic energy in the t-J model
\cite{wrobel,io,kotliar}. In this paper we set the frequency cut-off
at 1.25 eV and compare the results with the predictions of BCS
theory and CDMFT calculations based on the t-J model. In Fig.
\ref{fig7} we show a comparison between $W(T)$ for different samples
with different doping levels. One can clearly see that the onset of
superconductivity induces a positive change of the SW(0-1.25 eV) in
the underdoped sample and in the optimally doped one\cite{hajo}; in
the 77 K sample no superconductivity induced effect is detectable
for this frequency cut off and in the strongly overdoped sample we
observe a decrease of the low frequency spectral weight. In the
righthand panel of Fig. \ref{fig7} we also display the derivative of
the integrated SW as a function of temperature. The effect of the
superconducting transition is visible in the underdoped sample and
in the optimally doped sample as a peak in the derivative plot; no
effect is detectable in the overdoped 77 K sample, while in the
strongly overdoped sample a change in the derivative of the opposite
sign is observed.

\begin{figure}[ht]
    \centerline{\includegraphics[width=9cm,clip=true]{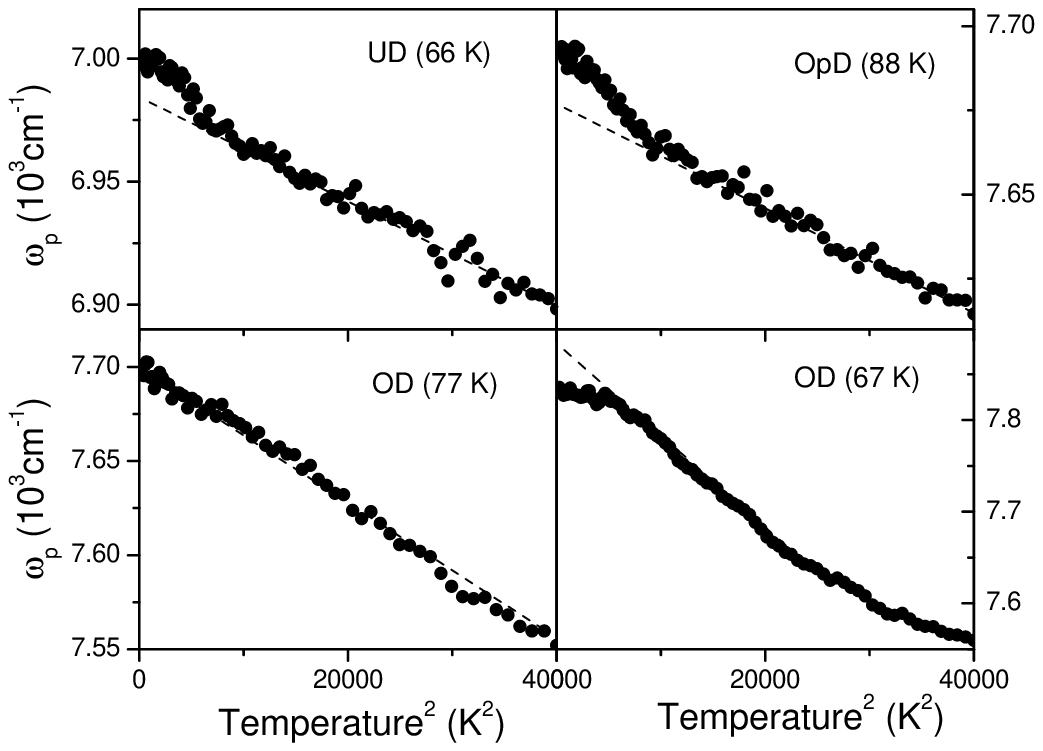}
    \includegraphics[width=9cm,clip=true]{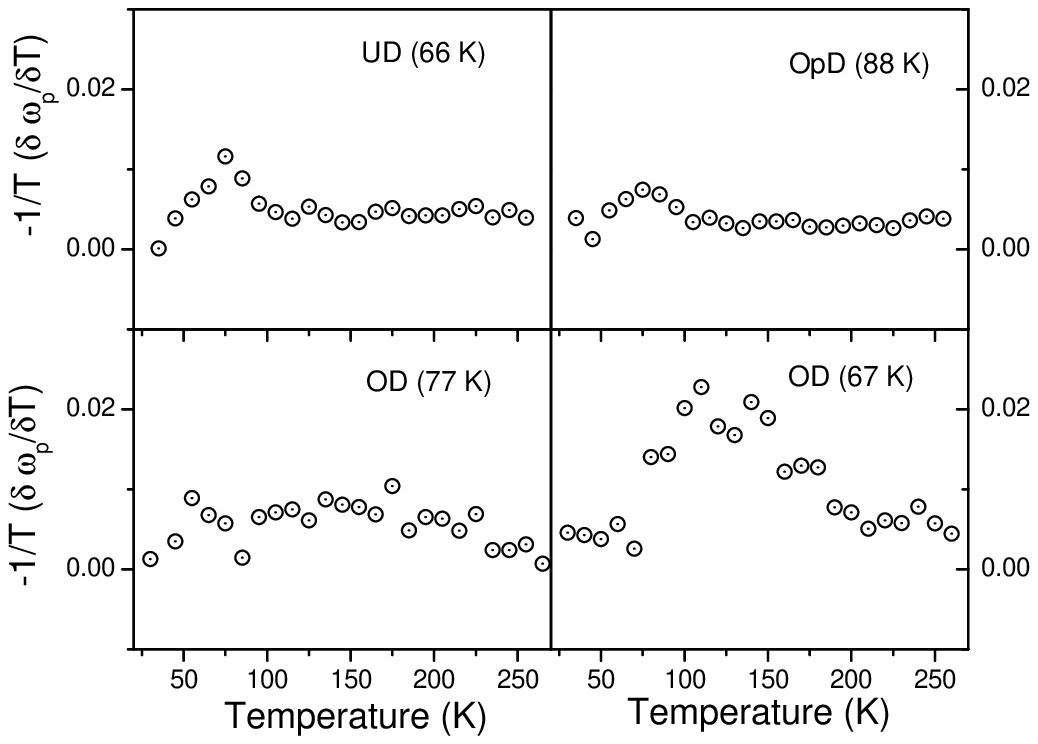}}
    \caption{Left panel: Screened plasma frequency as a function of temperature for different doping levels.
    Right panel:Derivative as a function of temperature, ($-\partial \omega_p/\partial T$), of the Screened plasma frequency for different doping levels.}
    \label{fig8}
 \end{figure}

The frequency $\omega_p^*$ for which $\epsilon_1(\omega_p^*)$ = 0
corresponds to the eigenfrequency of the longitudinal oscillations
of the free electrons for $k\rightarrow 0$.  $\omega_p^*$ can be
read off directly from the ellipsometric spectra, without any
data-processing. The temperature dependence of $\omega_p^*$ is
displayed in Fig. \ref{fig8}. The screened plasma-frequency has a
red shift due to the bound-charge polarizability, and the interband
transitions. Therefore its temperature dependence can be caused by a
change of the free carrier spectral weight, the dissipation, the
bound-charge screening, or a combination of those. This quantity and
its derivative as a function of temperature can clarify whether a
real superconductivity-induced change of the plasma frequency is
already visible in the raw experimental data. In view of the fact
that the value of $\omega_p^*$ is determined by several factors, and
not only the low frequency SW, it is clear that the SW still has to
be determined from the integral of Eq. \ref{equation1}. It is
perhaps interesting and encouraging to note, that in all cases which
we have studied up to date, the temperature dependences of $W(T)$
and $\omega_p^{*}(T)^2$ turned out to be very similar.

One can clearly see in the underdoped and in the optimally doped
sample that superconductivity induces a blue shift of the screened
plasma frequency. A corresponding peak is observed at $T_c$ in the
derivative plots. In the 77 K sample no effect is visible at $T_c$
while the 67 K sample shows a red shift of the screened plasma
frequency. The behavior of the screened plasma frequency also seems
to exclude the possibility that a narrowing with temperature of the
interband transitions around 1.5 eV is responsible for the observed
changes in the optical constants. If this would be the case then one
would expect the screened plasma frequency to exhibit a
superconductivity-induced shift in the same direction for all the
samples.

\subsection{Predictions for the spectral weight using the BCS model}
\begin{figure}[ht]
    \centerline{\includegraphics[width=8cm,clip=true]{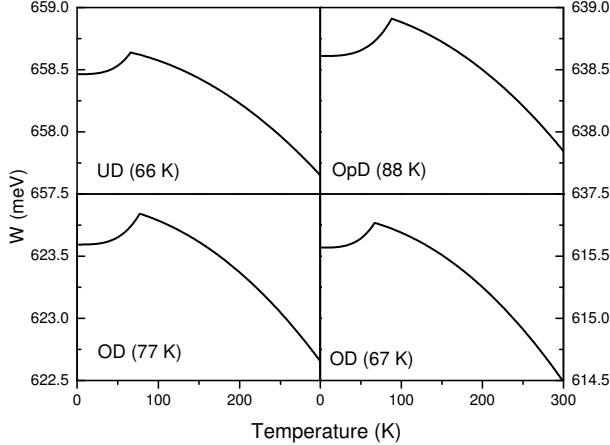}}
    \caption{Low frequency SW as a function of temperature calculated for the same doping levels experimentally measured.}
    \label{fig9}
 \end{figure}

In order to put the data into a theoretical perspective, we have
calculated $W(T)$ in the BCS model, using a tight-binding
parametrization of the energy-momentum dispersion of the normal
state. The parameters of the parametrization are taken from ARPES
data \cite{golden}. The details of this calculation are discussed in
the Appendix. Because in this parametrization both $t'$ and $t''$
are taken to be different from zero, the spectral weight is not
strictly proportional to the kinetic energy. Nonetheless for the
range of doping considered here, $W$ follows the same trend as the
actual kinetic energy, as has been pointed out previously by some of
us\cite{marel_hvar2002}. Results for the
$t-t^{\prime}-t^{\prime\prime}$ model are shown in Fig. \ref{fig9}.
We do wish to make a cautionary remark here, that a sign change as a
function of doping is not excluded {\em a priori} by the BCS model.
However, in the present case this possibility appears to be excluded
in view of the state of the art ARPES results for the
energy-momentum dispersion of the occupied electron bands. One can
see that for all considered doping levels, W decreases below $T_c$,
thus BCS calculations fail reproducing the temperature dependence in
the underdoped and optimally doped samples.

\subsection{Superconductivity induced transfer of spectral weight: experiment and cluster DMFT calculations}
\begin{figure}[ht]
    \centerline{\includegraphics[width=9cm,clip=true]{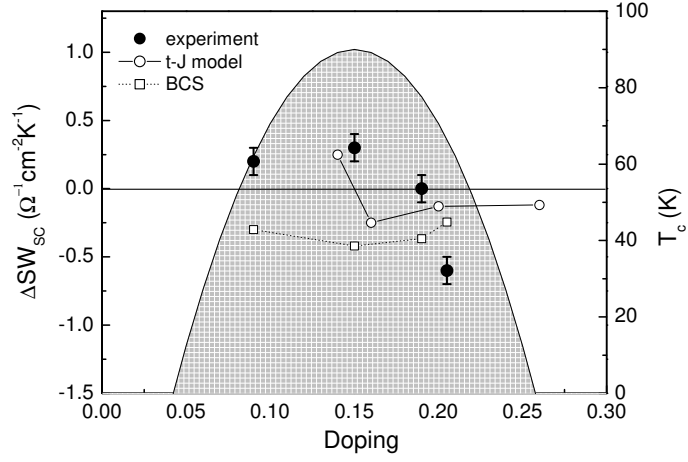}}
    \caption{Doping dependence of the superconductivity induced SW changes: experiment vs. theory. Two theoretical calculations are presented:
    d-wave BCS model and CDMFT calculations in the framework of the t-J model.}
    \label{ddq1}
 \end{figure}

\begin{figure}[ht]
    \centerline{\includegraphics[width=9cm,clip=true]{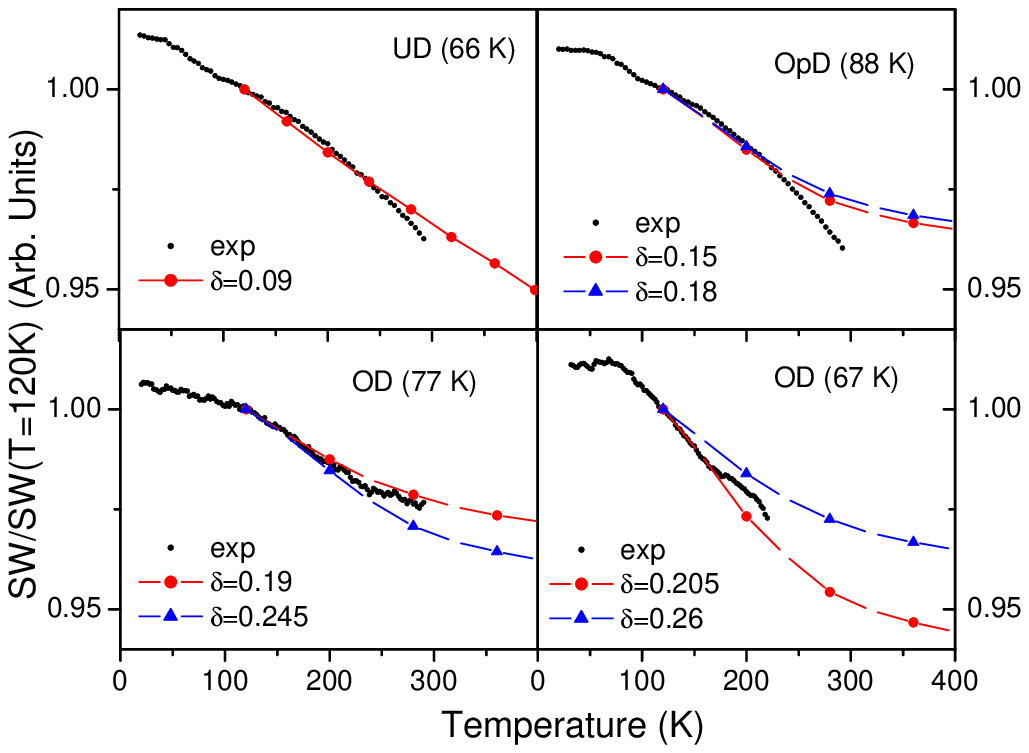}
    \includegraphics[width=9cm,clip=true]{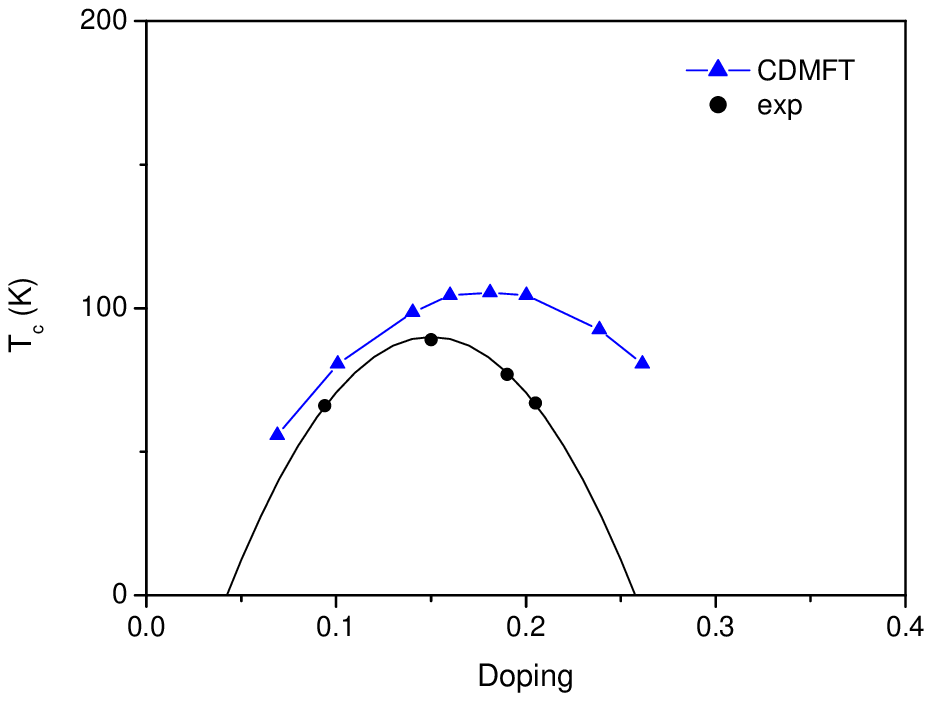}}
    \caption{Left panel: Comparison between the experimental and the theoretical $W(T)$ in the normal state for different doping levels.
    Right panel: comparison between the 'dome' as derived from theory and the experimental one.}
    \label{ddqns}
 \end{figure}

\begin{figure}[ht]
    \centerline{\includegraphics[width=10cm,clip=true]{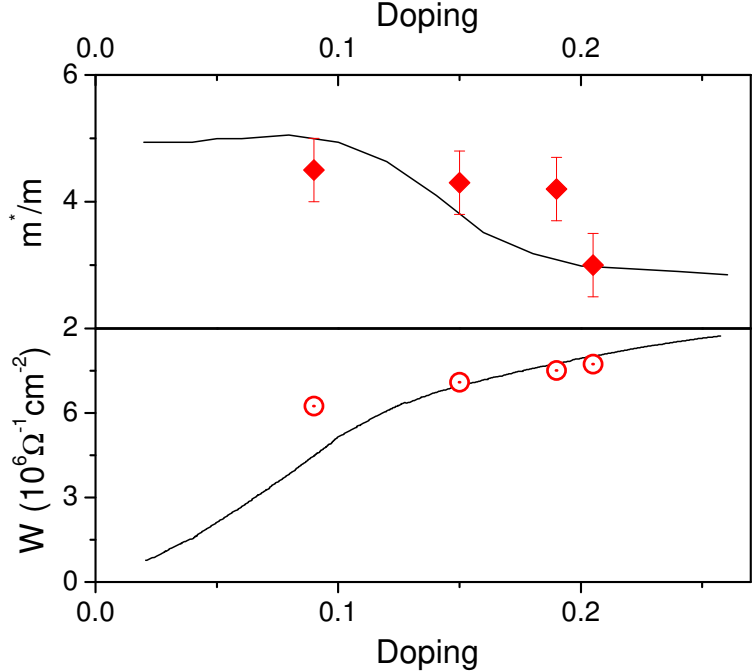}}
    \caption{Comparison between the calculated plasma frequency and effective mass and the experimental values.}
    \label{ddq2}
\end{figure}

In order to highlight the effect of varying the doping
concentration, we have extrapolated the temperature dependence in
the normal state of $W(\Omega_c,T)$ of each sample to zero
temperature, and measured it's departure from the same quantity in
the superconducting state, also extrapolated to T=0: $\Delta
SW_{sc}\equiv W(T=0)-W_n^{ext.}(T=0)$ In Fig. \ref{ddq1} the
experimentally derived quantities are displayed together with the
recent CDMFT calculations of the t-J\cite{kotliar} model and those
based on the BCS model explained in the previous subsection. While
the BCS-model provides the correct sign only for the strongly
overdoped case, the CDMFT calculations based on the t-J model are in
qualitative agreement with our data and the data in Ref
\onlinecite{deutscher}, insofar both the experimental result and the
CDMFT calculation give $\Delta SW_{sc}>0$ on the underdoped side of
the phase diagram, and both have a change of sign as a function of
doping when the doping level is increased toward the overdoped side.
The data and the theory differ in the exact doping level where the
sign change occurs. This discrepancy may result from the fact that
for the CDMFT calculations the values $t'=t''=0'$ were adopted. This
choice makes the shape of the Fermi surface noticeably different
from the experimentally known one, hence the corresponding
fine-tuning of the model parameters may improve the agreement with
the experimental data. This may also remedy the difference between
the calculated doping dependence of $T_c$ and the experimental one
(see righthand panel of Fig. \ref{ddqns}). We also show, in Fig.
\ref{ddq2}, the doping dependence of the plasma frequency and
effective mass compared to the CDMFT results. One can see that a
reasonable agreement is achieved for both quantities.

\subsection{Normal state trend of the spectral weight}
The persistence of the $T^2$ temperature dependence up to energies
much larger than what usually happens in normal metals has been
explained in the context of the Hubbard model \cite{calvani},
showing that electron-electron correlations are most likely
responsible for this effect. Indeed, experimentally we observe a
strong temperature dependence of the optical constants at energies
as high as 2 eV. In most of the temperature range, particularly for
the samples with a lower doping level, these temperature
dependencies are quadratic. Correspondingly, $W(T)$ also manifests a
quadratic temperature dependence. For sample OD67 the departure from
the quadratic behavior is substantial; the overall normal state
temperature dependence at this doping is also much stronger than in
the other samples. We compare, in Fig. \ref{ddqns}, the temperature
dependence of the SW with the predictions of the Hubbard model.

In Fig. \ref{ddqns} the experimental $W(T)$ is compared to the CDMFT
calculations for the same doping concentration. Since the $T_c$
obtained by CDMFT differs from the experimental one, (see Fig.
\ref{ddqns}) it might be more realistic to compare theory and
experiment for doping concentrations corresponding to the same
relative $T_c$'s. Therefore we also include in the comparison the
CDMFT calculation at higher doping level, at which $T_c/T_{c,max}$
corresponds to the experimental one (see the right panel of Fig.
\ref{ddqns}). We see that the experimental and calculated values of
$W(T)$ are in quantitative agreement for the temperature range where
they overlap. It is interesting in this connection, that the
curvature in the opposite direction, clearly present in all CDMFT
calculations, may actually be present in the experimental data, at
least for the highly doped samples. These observations clearly call
for an extension of the experimental studies to higher temperature
to verify whether a cross-over of the type of temperature dependence
of the spectral weight really exists, and to find out the doping
dependence of the cross-over temperature. The experimental data, as
mentioned before, show a rapid increase of the slope of the
temperature dependence above optimal doping. This behavior is
qualitatively reproduced by the CDMFT calculations.

One can calculate the normal state kinetic energy of the charge
carriers and its temperature dependence starting from the
tight-binding dispersion relation neglecting the correlation
effects. In this context, one can find a stronger temperature
dependence of the normal state SW when the chemical potential
approaches the van Hove singularity. Extrapolating the experimental
bandstructure beyond x=0.22 we estimate that this would happen at a
doping level as high as 0.4 in Bi2212 \cite{golden}. This offers an
alternative scenario for the normal state temperature dependence,
although the role of the van Hove singularity has to be explored in
further detail. We also point out that as a result of crossing this
singularity one can get a SW increase in the superconducting state
within the BCS model. In the CDMFT calculations presented in Fig.
\ref{ddq1} the SW temperature dependence in the normal state is a
pure correlation effect, since in this calculation the van Hove
singularity is located at exactly half filling, far away from the
experimentally considered doping levels.

\section{CONCLUSIONS}

In conclusion, we have studied the doping dependence of the optical
spectral weight redistribution in single crystals of Bi2212, ranging
from the underdoped regime, $T_c$ = 66 K to the overdoped regime,
$T_c$ = 67 K. The low frequency SW increases when the system becomes
superconducting in the underdoped region of the phase diagram, while
it shows no changes in the overdoped sample $T_c$ = 77 K and
decreases in the $T_c$ = 67 K sample. We compared these results with
BCS calculations and CDMFT calculations based on the t-J model. We
show that the latter are in good qualitative agreement with the data
both in the normal and superconducting state, suggesting that the
redistribution of the optical spectral weight in cuprates
superconductors is ruled by electron-electron correlations effects.

\section*{ACKNOWLEDGMENTS}

We are grateful to T. Timusk, N. Bontemps, A.F. Santander-Syro, J.
Orenstein, and C. Bernhard for stimulating discussions. This work
was supported by the Swiss National Science Foundation through the
National Center of Competence in Research "Materials with Novel
Electronic Properties-MaNEP".

\section{APPENDIX}

The pair formation in a superconductor can be described by a
spatial correlation function $g(r)$ which has a zero average in
the normal state and a finite average in the superconducting
state. Without entering into the details of the mechanism itself
responsible for the attractive interaction between electrons, one
can assume that an attractive potential V(r) favors a state with
enhanced correlations in the superconducting state. In the
superconducting state the interaction energy differs from the
normal state by:
\begin{equation}
<H^i>_s - <H^i>_n=\int dr^3g(r)V(r)
\end{equation}
With some manipulations one can relate the correlation function to
the gap-function and the single particle energy. The result is a BCS
equation for the order parameter, with a potential which can be
chosen to favor pairing with d-wave symmetry. The simplest approach
is to use a simple separable potential which leads to an order
parameter of the form, $\Delta_k = \Delta_0(T) (\cos{k_x} -
\cos{k_y})/2$. The temperature dependence of $\Delta_0(T)$ can then
be solved as in regular BCS theory. We have done this for a variety
of parameters \cite{frank}, and find that $\Delta_0(T)/\Delta_0(0) =
\sqrt{1 - (t^4+t^3)/2}$ gives a very accurate result (for either
s-wave or d-wave symmetry), where $t \equiv T/T_c$. Then, for
simplicity, we adopt the weak coupling result that $\Delta_0(0) =
2.1 k_B T_c$. Finally, even in the normal state, the chemical
potential is in principle a function of temperature (to maintain the
same number density); this is computed by solving the number
equation, $n = ({2 \over N})\sum_k n_k$, where
$$
n_k = 1/2 - (\epsilon_k - \mu) {[1 - 2f(E_k)] \over 2E_k}
$$
where $E_k \equiv \sqrt{(\epsilon_k - \mu)^2 + \Delta_k^2}$ at each
temperature for $\mu$ for a fixed doping. Once these parameters are
determined, one can calculate the spectral weight sum, $W$, for a
given band structure. We use:
$$
\epsilon_k = -2t*(cos(k_x) + cos(k_y)) + 4t^\prime*cos(k_x)cos(k_y)
- 2t^{"}*(cos(2k_x) + cos(2k_y)).
$$
In this expression $\delta$ is the hole doping, and $\Delta_0$ is
the gap value calculated as $2.1K_BT_c$, t = 0.4 eV, t$^\prime$ =
0.09 eV and t$^{\prime\prime}$ = 0.045 eV. The dispersion is taken
from ARPES measurements \cite{golden}; for simplicity we have left
out the bi-layer splitting and the constant. The spectral weight sum
is given by
$$
W = \sum_k {\partial_k^2 \epsilon_k \over \partial k_x^2} n_k.
$$
Results are plotted in Fig. 6 for the doping levels of the samples
used in the experiments. These calculations clearly show that BCS
theory predicts a lowering of the spectral weight sum in the
superconducting phase; this is in disagreement with the experimental
results in the underdoped and optimally doped samples. Moreover,
there is no indication of a change of sign of the
superconductivity-induced SW changes in this doping interval within
the BCS formalism. Note, however, that preliminary calculations
indicate that the van Hove singularity can play a role at much
higher doping levels (not realized, experimentally), and that in
theory a sign change in the anomaly can occur even within BCS
theory.

\end{document}